\documentclass[12pt]{iopart}
\pdfoutput=1
\usepackage{iopams}
\usepackage{graphicx}

\def\re{\mathop{\rm Re}\nolimits}
\def\im{\mathop{\rm Im}\nolimits}

\def\Tr{\mathop{\rm Tr}\nolimits}

\newtheorem{theo}{Theorem}

\newtheorem{prop}{Proposition}
\begin{document}
\title[Determination of $S$-curves]{Determination of $S$-curves with applications
                                                       to the theory of nonhermitian orthogonal polynomials}
\author{ Gabriel \'Alvarez$^1$, Luis Mart\'{\i}nez Alonso$^1$ and Elena Medina$^2$}
\address{$^1$ Departamento de F\'{\i}sica Te\'orica II,
                        Facultad de Ciencias F\'{\i}sicas,
                        Universidad Complutense,
                        28040 Madrid, Spain}
\address{$^2$ Departamento de Matem\'aticas,
                        Facultad de Ciencias,
                        Universidad de C\'adiz,
                        11510 Puerto Real, C\'adiz, Spain}
\begin{abstract}
This paper deals with the determination of the $S$-curves in the theory of non-hermitian orthogonal
polynomials with respect to exponential weights along suitable paths in the complex plane.
It is known that the corresponding complex equilibrium potential can be written as a combination of Abelian
integrals on a suitable Riemann surface whose branch points can be taken as the main parameters
of the problem. Equations for these branch points can be written in terms of periods
of Abelian differentials and are known in several equivalent forms. We select one of these forms and
use a combination of analytic an numerical methods to investigate the phase structure of asymptotic
zero densities of orthogonal polynomials and of asymptotic eigenvalue densities of random
matrix models.  As an application we give a complete description of the phases and critical
processes of the standard cubic model.
\end{abstract}
\maketitle
\section{Introduction}
The present paper elaborates on the notion of $S$-curve of Stahl~\cite{ST85a,ST85b,ST86a,ST86b}
and of Gonchar and Rakhmanov~\cite{GO89,RA11}. Among the many applications of
$S$-curves (see for instance section~6.3 of~\cite{MA11} and references therein),
we pay special attention to the theory of nonhermitian orthogonal polynomials  $p_n(z)=z^n+\cdots$  
\begin{equation}
    \label{pol1}
    \int_{\Gamma}p_n(z) z^k \rme^{-n W(z)}  \rmd  z=0,\quad k=0,\ldots,n-1.
\end{equation}
The classical theory of orthogonal polynomials corresponds to the hermitian case, in which the integration
path $\Gamma$ is typically a real interval and the weight is a positive real function on $\Gamma$.
But more recently the nonhermitian case, in which $\Gamma$ can be a more general curve in the complex
plane and the weight can be a complex function, has received much attention.
In the mathematical literature these polynomials first appeared as denominators
of Pad\'e and other types of rational approximants~\cite{ST85a,ST85b,ST86a,ST86b}, but the corresponding
theory quickly developed and found applications into such fields as the Riemann-Hilbert approach to strong
asymptotics, random matrix theory~\cite{DE99b,BL99,BL03,BL08,BE09,BE11} and, consequently, in the study
of dualities between supersymmetric gauge theories and string models~\cite{CA01,DI02,DI022,HE07,MAR10}.

More concretely, our aim is to apply the general theory of $S$-curves as developed in~\cite{GO89,RA11,MA11}
to study the asymptotic distribution of zeros of orthogonal polynomials and the phase structure
of the asymptotic distribution of eigenvalues  as $n\rightarrow\infty$ of random matrix problems of the
form~\cite{DA91,DA93,DE99,FE04,LA03}
 \begin{equation}
     \label{mm0}
    Z_n=\int  \rmd  M \rme^{-n  {\rm Tr}  W(M)},
\end{equation}
where the eigenvalues of the $n\times n$ matrices $M$ are constrained to lie on $\Gamma$.  
 
Throughout our discussion we assume that  $W(z)$ is a complex polynomial and $\Gamma$ is a simple  
analytic curve connecting two different convergence sectors ($\re W(z)>0$) at infinity of~(\ref{pol1}). 
A fundamental result of Gonchar and Rakhmanov~\cite{GO89} asserts that if $\Gamma$ is an $S$-curve, then
the  asymptotic zero distribution of $p_n(z)$  exists and is given by the equilibrium charge density~\cite{SA97}  that
minimizes the electrostatic energy (among normalized charge densities supported on the curve $\Gamma$)
in the presence of the external electrostatic potential $V(z)=\re W(z)$. Note that the integral~(\ref{pol1})
is invariant under deformations of the curve $\Gamma$ into curves in the same homology class and connecting
the same two convergence  sectors at infinity.  This freedom to deform $\Gamma$ means that only for special
choices of $\Gamma$ the asymptotic zero distribution has support on $\Gamma$.  According to recent
results by Rakhmanov~\cite{RA11}, given a family of orthogonal polynomials of the form~(\ref{pol1})
we can always deform $\Gamma$ into an appropriate  $S$-curve.
 
We use an analytic scheme, to be implemented in general with the help of numerical analysis,
based on the study of certain algebraic curves which arise as a direct consequence of the
$S$-property~\cite{GO89,RA11,MA11}. These \emph{spectral curves} have the form
 \begin{equation}
    \label{eq:c0}
    y^2 = W'(z)^2 + f(z),
\end{equation}
where $f(z)$ is a  polynomial such that $ \deg  f= \deg  W-2$. The main parameters that determine
the $S$-curves and the associated equilibrium densities are the branch points of $y(z)$, which turn
out to be the endpoints of the (in general, several disjoint) arcs (cuts) that support the equilibrium density.
Systems of equations for these branch points can be formulated in terms of period integrals of $y(z)$
and are known in several equivalent forms. We select one of these forms that in the Hermitian case
reduces to the system of equations derived in~\cite{AL10}. The corresponding cuts are characterized
as Stokes lines of the polynomial $y(z)^2$ or, equivalently, as trajectories of the quadratic differential
$y(z)^2 ( \rmd z)^2$. At this point we use numerical analysis not only to solve the equations for the
cut endpoints but also to analyze the existence of cuts satisfying the $S$-property. 

Recently Bertola and Mo~\cite{BE09} and Bertola~\cite{BE11} have used the notion of Boutroux
curves  to characterize the support of the asymptotic distribution of zeros of families of nonhermitian
orthogonal polynomials.  Both the calculations of the present paper and the approach of~\cite{BE09,BE11}
do not rely on the minimization of a functional but on the characterization of spectral curves~(\ref{eq:c0})
with appropriate cuts.  This characterization is formulated in~\cite{BE09,BE11} in terms of admissible
Boutroux curves which are determined from certain combinatorial and metric data in the space of
polynomials $p(z)$.  It can be proved that  the branch cut structure of admissible Boutroux curves
consists of arcs satisfying the $S$-property and, consequently,  the method  of~\cite{BE09,BE11}
can also be  applied to  characterize $S$-curves.  However, as we explained in the previous paragraph,
our calculations are based on an explicit system of equations for the cut endpoints. In contrast, 
the generation of nontrivial explicit examples in~\cite{BE09,BE11} amounts to imposing directly period
conditions by means of a numerical algorithm involving the minimization of a functional that vanishes
precisely for admissible Boutroux curves.

The paper is organized as follows. In section~\ref{sec:zdop} we review the basic results on equilibrium
densities of electrostatic models under the action of external fields. Then we introduce the notions of $S$-curve
and $S$-property, and discuss their relevance to characterize asymptotic zero densities of orthogonal polynomials.
To obtain an equivalent but computationally more efficient formulation of the $S$-property it is convenient
to introduce the complex counterpart of the  electrostatic potential. This formulation leads naturally to the notion
of spectral curve. In section~\ref{sec:ced} we recall the theoretical background to construct equilibrium
densities on $S$-curves  for a given polynomial $W(z)$ and use the theory of Abelian differentials
in Riemann surfaces to derive a system of equations for the cut endpoints. We also discuss the
characterization of cuts as Stokes lines and the process of embedding the cuts into $S$-curves.
In section~\ref{sec:tcm} we apply the former results to perform a complete analysis of the cubic model
\begin{equation}
    W(z) = \frac{z}{3} - t z,
\end{equation}
with a varying complex coefficient $t$. We determine $S$-curves and equilibrium densities for the two possible
cases corresponding to equilibrium densities supported on one or two disjoint arcs. Our analysis combines theoretical
properties with numerical calculations and allows us to characterize critical processes of  merging, splitting, birth
and death at a distance of cuts. As a consequence we describe the phase structure of the corresponding
families of orthogonal polynomials on different paths $\Gamma$. The consistency of our results is checked by
superimposing the cuts and the zeros of the corresponding orthogonal polynomials $p_{n}(z)$ with degree
$n=24$. Thus we find a complete agreement with the  Gonchar-Rakhmanov Theorem~\cite{GO89}
(Theorem~1 below). Finally,  in section~\ref{sec:gcr}  we briefly discuss a generalization
of the $S$-property which  arises in the study of dualities between supersymmetric gauge theories and string
models on local Calabi-Yau manifolds.  Some technical aspects of the theoretical discussion are treated
in appendix~A.
\section{Zero densities of orthogonal polynomials\label{sec:zdop}}
According to  the general theory of logarithmic potentials with external fields~\cite{SA97}, given an
analytic curve $\Gamma$ in the complex plane and a real-valued external potential $V(z)$, there exists
a unique charge density $\rho(z)$ that minimizes the total electrostatic energy 
\begin{equation}
	\label{min2}
 	\mathcal{E}[\rho]
	=
	\int_{\Gamma} |\rmd z| \rho(z)  V(z)
	-
	\int_{\Gamma} |\rmd z| \int_{\Gamma} |\rmd z'| \log |z-z'| \rho(z) \rho(z')
\end{equation}
among all positive densities supported on $\Gamma$ such that
\begin{equation}
	\label{norm}
	\int_{\Gamma} |\rmd z| \rho(z) = 1.
\end{equation}
This density $\rho(z)$ is called the \emph{equilibrium density}, and its  support $\gamma$ is a finite union
of disjoint analytic arcs $\gamma_i$ (cuts) contained in $\Gamma$:
\begin{equation}
	\label{cuts}
	\gamma = \gamma_1\cup\gamma_2\cup \cdots\cup \gamma_s\subset\Gamma.
\end{equation}
In terms of the total electrostatic potential 
\begin{equation}
	\label{loge}
	U(z) = V(z) - 2 \int_{\Gamma} |\rmd z'| \rho(z') \log |z-z'|,
\end{equation}
the equilibrium density is characterized by the existence of a real constant $l$ such that
\begin{equation}
	\label{cm}
	U(z) = l,\quad z\in \gamma,
\end{equation}
\begin{equation}
	\label{cm1}
	U(z) \geq l,\quad z\in \Gamma-\gamma.
\end{equation}

The property that relates this minimization problem to the asymptotic zero density of orthogonal polynomials
is called the \emph{S-property}, and was singled out by Stahl~\cite{ST85a,ST85b,ST86a,ST86b},
elaborated by Gonchar and Rakhmanov~\cite{GO84,GO89}, and more recently extended by
Mart\'{\i}nez-Finkelshtein and Rakhmanov~\cite{MA11}.

A curve  $\Gamma$ is said to be an \emph{$S$-curve}
with respect to the external field $V(z)$ if for every interior point $z$ of the support $\gamma$ of the equilibrium
density the total potential~(\ref{loge}) satisfies
\begin{equation}
	\label{s0}
	\frac{\partial U(z)}{\partial n_+} =  \frac{\partial U(z)}{\partial n_-},
\end{equation}
where $n_{\pm}$ denote the two normal vectors to $\gamma$ at $z$ pointing in the opposite directions.
In this case it is said that $\gamma$ satisfies the $S$-property.
The condition~(\ref{s0}) means that  the electric fields
at each side are opposite, $\mathbf{E}_{+}=-\mathbf{E}_{-}$. 
\subsection{Orthogonal polynomials and  $S$-curves\label{sec:eds}}
Let  $\{p_n(z)\}_{n\geq 1}$  be a family  of monic orthogonal polynomials on a curve  $\Gamma$
with respect to an exponential weight $\exp(-n W(z))$,
\begin{equation}
	\label{pol1b}
        \int_{\Gamma} p_n(z) z^k \rme^{-n W(z)} \rmd  z=0,\quad k=0,\ldots,n-1.
\end{equation}
Here and henceforth we assume that $W(z)$ is a complex polynomial of degree $N+1$
\begin{equation}
W(z)=\sum_{k=1}^{N+1} t_k\, z^k,
\end{equation}
and  that $\Gamma$ is an oriented  simple analytic curve which as $z\rightarrow \infty$ connects two different
sectors  of convergence of~(\ref{pol1b}).  The notion of $S$-curve is crucial in the analysis of the limit
as $n\rightarrow \infty$ of the  zero density  $\frac{1}{n}(\delta(z-c_1)+\cdots+\delta(z-c_n))$ of $p_n(z)$.
The following Theorem~(see~\cite{GO89}, section~3) states the close relation between the asymptotic
zero distribution of orthogonal polynomials and the equilibrium densities on $S$-curves:
\begin{theo}
Let  $\{p_n(z)\}_{n\geq 1}$  be a family  of orthogonal polynomials on a curve  $\Gamma$ with respect
to an exponential weight $\exp(-n W(z))$.   If $\Gamma$ is an  $S$-curve with respect to the external
potential $V(z)=\re W(z)$, then  the  equilibrium density on  $\Gamma$ is  the  weak limit
as $n\rightarrow \infty$ of the zero density  of $p_n(z)$.
\end{theo}

It often occurs in the applications that the orthogonal polynomials $p_n(z)$ are initially defined
on a curve $\Gamma$ which is not an $S$-curve. This problem raises the question of the
existence of an $S$-curve in the same homology class of $\Gamma$ connecting the same pair
of convergence sectors at infinity (and therefore defining the same family of orthogonal polynomials).
This question has been recently solved in the affirmative by Rakhmanov~(see~\cite{RA11}, section~5.3).
Note also that although this $S$-curve is not unique, the associated equilibrium density is certainly unique.
\subsection{Matrix models }
Equilibrium densities on $S$-curves are also expected to describe the asymptotic eigenvalue distribution
as $n\rightarrow \infty$ of random matrix models  with partition function~(\ref{mm0}). According to
Heine's formula~\cite{DE99}, the polynomials~(\ref{pol1}) are the expectation values of the
characteristic polynomials of the matrices $M$ of the ensemble,
\begin{equation}
	p_n(z) = \frac{1}{Z_n} \int  \rmd  M \det(z-M) \rme^{-n \Tr W(M)}.
\end{equation}
In terms of the zeros $\{c_i\}_{i=1}^n$ of $p_n(z)$ and of the eigenvalues $\{\lambda_i\}_{i=1}^n$  of $M$,
this result means that the expectation value of the function $\prod_{i=1}^n (z-\lambda_i)$ is the function
$\prod_{i=1}^n (z-c_i)$. Therefore it is natural to conjecture that the asymptotic distributions
of zeros of $p_n(z)$ and of eigenvalues of $M$ coincide. This conjecture has been rigorously proved
in the hermitian case, i.e., when $\Gamma=\mathbb{R}$ and the polynomial $W(z)$
has real coefficients~\cite{DE99b,DE99}, and indeed orthogonal polynomials are a widely used tool
in many aspects of hermitian random matrix theory (for some recent applications see~\cite{NA11,AL11}).
\subsection{Spectral curves}
The $S$-property can be formulated  in a more convenient form to our goals using a complex counterpart
of the electrostatic potential~(\ref{loge}). Thus, we define 
\begin{equation}
    \label{comp0}
    \mathcal{U}(z) = W(z) - (g(z_+)+g(z_-)),
\end{equation}
where $g(z)$ is the analytic function  in $\mathbb{C}\setminus \Gamma$ given by
\begin{equation}
    \label{ge}
    g(z) = \int_{\gamma}| \rmd z'| \rho(z')\log(z-z').
\end{equation}
Here we assume that the logarithmic branch is taken in such a way that for every $z'\in \Gamma$
the function $\log(z-z')$ is an analytic  function of $z$ in $\mathbb{C}$ minus the semi-infinite arc of $\Gamma$
ending at $z'$. As usual $g(z_+)$ and $g(z_-)$ denote the limits of the function $g(z')$ as $z'$ tends to $z$
from the left and from the right of the oriented curve $\Gamma$ respectively.

It is clear  that
\begin{equation}
 \re \mathcal{U}(z) = U(z),
\end{equation}
and therefore the equilibrium condition~(\ref{cm}) can be rewritten as
\begin{equation}
    \label{mina1}
     \re \mathcal{U}(z) = l, \quad z\in \gamma.
\end{equation}
Furthermore, it follows from the Cauchy-Riemann equations that  the $S$-property~(\ref{s0})
is verified if and only if the imaginary part of $\mathcal{U}(z)$ is constant on each arc $\gamma_j$
of $\gamma$ (usually stated as ``locally constant on $\gamma$'')~\cite{MA11,DE10}:
\begin{equation}
    \label{s20}
    \im  \mathcal{U}(z)=m_j, \quad  z\in \gamma_j,\quad j=1,\ldots, s.
\end{equation}
Note that, in essence, the $S$-property embodies the possibility of analytically continuing
the derivative of the complex equilibrium potential through the support. In some physical
applications~\cite{CA01,IT03,IT03b} the values $L_j = l + \rmi m_j$ are especially relevant,
and equations~(\ref{mina1}) and~(\ref{s20}) are (trivially) restated by saying that $\Gamma$
is an $S$-curve if and only if the complex potential $\mathcal{U}(z)$ is locally constant on $\gamma$
    \begin{equation}
        \label{s2}
        \mathcal{U}(z)=L_j, \quad  z\in \gamma_j,\quad j=1,\ldots, s,
    \end{equation}
    and the constants $L_j$ have the same real part
    \begin{equation}
        \label{s22}
         \re L_1=\cdots = \re L_s.
    \end{equation}
 
Next we will see how equations~(\ref{s2}) lead to the notion of spectral curve.
(In section~\ref{sec:eqendpts} we will see that equations~(\ref{s22})
are essential to formulate the system of equations for the cut endpoints in the multicut case.)
In fact, condition~(\ref{s2}) can be rewritten in a form especially suited for practical applications
in terms of a new function $y(z)$ defined by
\begin{equation}
    \label{y}
    y(z) = W'(z)-2 g'(z)
           = W'(z)-2 \int_{\gamma}| \rmd z'| \frac{\rho(z')}{z-z'}, \quad z\in\mathbb{C}\setminus \Gamma.
\end{equation}
\begin{prop}
    The complex potential $\mathcal{U}(z)$ is locally constant on $\gamma$ if and only if the square of $y(z)$
    is a polynomial  of the form
    \begin{equation}
        \label{c1b}
        y(z)^2 = W'(z)^2 + f(z),
    \end{equation}
    where $f(z)$ is a polynomial of degree $ \deg  f= \deg  W-2$.
\end{prop}
\emph{Proof.}
Condition~(\ref{s2}) is equivalent to
\begin{equation}
    \label{s3}
    W'(z) - ( g'(z_+) + g'(z_-) )=0,\quad z\in \gamma,
\end{equation}
where
\begin{equation}
    \label{gp}
    g'(z) = \int_{\gamma}| \rmd z'| \frac{\rho(z')}{z-z'},
\end{equation}
and therefore~(\ref{s3}) can be written as
\begin{equation}
    \label{rh}
    y(z_+) = -y(z_-),\quad z\in \gamma.
\end{equation}
The function $y(z)$ is analytic in $\mathbb{C}\setminus\gamma$ and, due to~(\ref{rh}),
its square is continuous on $\gamma$. Hence  $y^2(z)$ is analytic in the whole $\mathbb{C}$.
Furthermore, since
\begin{equation}
    \label{gpp}
    g'(z) = \frac{1}{z} + \mathcal{O}\left(\frac{1}{z^2}\right),\quad z\rightarrow \infty,
\end{equation}
we have that
\begin{equation}
    \label{asy}
    y(z) = W'(z)-\frac{2}{z}+\mathcal{O}\left(\frac{1}{z^2}\right),\quad z\rightarrow \infty,
\end{equation}
and Liouville's theorem implies that $y^2(z)$ is a polynomial of degree $2 N$ with $N= \deg  W-1$.
Therefore, we have
\begin{equation}
    \label{sc1}
    y^2=W'(z)^2+f(z)
\end{equation}
where
\begin{equation}
    \label{sc2}
    f(z)=y^2(z)-W'(z)^2=4 g'(z) (g'(z)-W'(z))
\end{equation}
is a polynomial of degree $N-1$. Reciprocally, given $\rho(z)$, if the function~(\ref{y}) is such that
its square is a polynomial then it satisfies~(\ref{rh}) and consequently~(\ref{s3}).

Equation~(\ref{c1b}) defines an algebraic curve referred to as a \emph{spectral curve}, which determines
the equilibrium charge density via~(\ref{y}),
\begin{equation}
    \label{mes}
    \rho(z) | \rmd z| = y(z_+) \frac{ \rmd z}{2 \pi\rmi}
                             = -y(z_-) \frac{ \rmd z}{2 \pi\rmi},\quad z\in\gamma,
\end{equation}
where $\gamma$ has the orientation inherited by the orientation of  $\Gamma$.
\subsection{The hermitian case}
Hermitian families of orthogonal polynomials correspond to $\Gamma=\mathbb{R}$ and $W(z)$ with real coefficients.
In this hermitian case it is clear that the real line $\Gamma=\mathbb{R}$ is an $S$-curve, because taking
$\log(z-z')$ as the principal branch of the logarithm we have
\begin{equation}
  \log(z_+-z')+\log(z_--z')=2\log |z-z'|,\quad z,z'\in \mathbb{R}.
\end{equation}
Hence~(\ref{s20}) holds because
\begin{equation}
    \im\mathcal{U}(x)= \im\left[W(x)-\left(g(x_+)+g(x_-)\right)\right]=0,\quad x\in \gamma\subset \mathbb{R}.
\end{equation}
There is a well stablished theory for characterizing the asymptotic distribution of zeros for hermitian orthogonal
polynomials and the asymptotic distribution of eigenvalues for hermitian matrix
models~\cite{DE99b, BL99,BL03,BL08,DE99}. In particular, a method of  analysis of the phase structure
and critical processes  for multicut hermitian matrix models was recently presented in~\cite{AL10}. 
\section{Construction of equilibrium densities on $S$-curves\label{sec:ced}}
In this section we discuss the theoretical background underlying the determination of $S$-curves.
Using Proposition~1, we begin by looking for spectral curves~(\ref{c1b}) where $\deg  f = \deg  W-2 = N-1$.
Obviously, the number $s$ of possible cuts for a fixed $W(z)$ is at most $N$.
We assume for simplicity that $y^2(z)$ has only simple or double roots. The simple roots will be denoted
by $\{a_j^{\pm}\}_{j=1}^{s}$ and the double roots by $\{\alpha_l\}_{l=1}^{r}$. The simple
roots $a_j^{\pm}$ will be the endpoints of the cut $\gamma_j$, and therefore $r+s=N$.

To determine the branch of $y(z)$ that verifies~(\ref{asy}) in the $s$-cut case, we write $y(z)$ in the form
 \begin{equation}
    \label{y1}
    y(z) = h(z) w(z),
\end{equation}
\begin{equation}
    \label{y2}
    h(z) =
          \prod_{l=1}^{r} (z-\alpha_l),
    \quad
    w(z) =  \sqrt{\prod_{m=1}^{s} (z-a_m^{-})(z-a_m^{+})},
    \end{equation}
and take the branch of $w(z)$  such that
\begin{equation}
    w(z)\sim z^{s},\quad  z\rightarrow\infty.
\end{equation}
The factor $h(z)$ in~(\ref{y1}) is then given by
\begin{equation}
	\label{ache}
	h(z) = \left(\frac{W'(z)}{w(z)}\right)_{\oplus},
\end{equation}
where $\oplus$ stands for the sum of the nonnegative powers of the Laurent series at infinity.
Hence the function $y(z)$ is completely determined by its branch points $\{a_j^{\pm}\}_{j=1}^{s}$,
and satisfies 
\begin{equation}
	y(z) = W'(z)+\mathcal{O}\left(\frac{1}{z}\right),\quad z\rightarrow \infty.
\end{equation}
Our first task is to find a system of equations for the cut endpoints $\{a_j^{\pm}\}_{j=1}^{s}$.  
\subsection{Equations for the cut endpoints\label{sec:eqendpts}}
We use the theory of Abelian differentials in Riemann surfaces to find a system of equations satisfied by the
cut endpoints (see Appendix~A for definitions and notations). Let us denote by $M$ the hyperelliptic Riemann surface
associated to the curve
\begin{equation}
  \label{1.1}
  w^2=\prod_{m=1}^{s} (z-a_m^{-})(z-a_m^{+}).
\end{equation}
We introduce  the meromorphic differential  $\mathrm{y}(z)\rmd z$ in $M$, where $\mathrm{y}(z)$
is the extension of the function~(\ref{y1}) to the Riemann surface $M$ in terms of two branches
of $\mathrm{y}(z)$ in $M$ given by $y_1(z)=-y_2(z)=y(z)$. The asymptotic condition~(\ref{asy}) implies
\begin{equation}
  \label{1.4}
\mathrm{y}(z)\rmd z   = \left\{\begin{array}{ll}
                      \displaystyle\left(W'(z)-\frac{2}{z}+\mathcal{O}(z^{-2})\right)\rmd z,
                      \quad\mbox{as $z\rightarrow \infty_1$},\\
                      \displaystyle\left(-W'(z)+\frac{2}{z}+\mathcal{O}(z^{-2})\right)\rmd z,
                      \quad \mbox{as $z\rightarrow \infty_2$}.
                     \end{array}\right.
\end{equation}
Since the only poles of $\mathrm{y}(z)\rmd z$ are at $\infty_1$ and $\infty_2$, equation~(\ref{1.4}) shows that
(with $t_0=-1$)
\begin{equation}
	\frac{1}{2}\left(\mathrm{y}(z)+W'(z)\right)\rmd z-\sum_{n=0}^{N+1} t_n \rmd \Omega_n
\end{equation}
is a first kind Abelian differential in $M$. Hence it admits a decomposition in the canonical basis
\begin{equation}
	\label{dba}
	\frac{1}{2}\left(\mathrm{y}(z)+W'(z)\right)\rmd z-\sum_{n=0}^{N+1} t_n \rmd \Omega_n
	=
	\sum_{j=1}^{s-1}\lambda_j  \rmd \varphi_j,
\end{equation}
for some complex coefficients $\lambda_j\in \mathbb{C}$. Thus, we may write
\begin{equation}
	\label{idy}
	\mathrm{y}(z) \rmd z
	=
	-W'(z) \rmd z+2 \sum_{j=1}^{s-1} \lambda_j \rmd \varphi_j +2 \sum_{n=0}^{N+1} t_n \rmd \Omega_n.
\end{equation}

Let us now denote by $\gamma_j$ a set of oriented cuts  joining the pairs  $a_j^-$ and $a_j^+$ of the
function~(\ref{y1}), and by $z_j$ an arbitrary point in $\gamma_j$. The $A$-periods of
the differential $\mathrm{y}(z)\rmd z$ can be written as
\begin{equation}
	A_j(\mathrm{y}(z)\rmd z)
	=
	\int_{z_{j}+}^{z_{(j+1)}+} y_1(z) \rmd z
	+
	\int_{z_{(j+1)}-}^{z_{j}-} y_2(z) \rmd z.
\end{equation}
Since $y_1(z)=-y_2(z)=y(z)$, we have that $y_2(z_-)=-y_2(z_+)=y_1(z_+)=-y_1(z_-)$.
Hence from~(\ref{y}) and~(\ref{s22}) we get
\begin{eqnarray}
	\label{as}
	\fl
	\nonumber A_j(\mathrm{y}(z)\rmd z)
	= 2 \left[W(z_{j+1})-\left(g(z_{(j+1)+})+g(z_{(j+1)-})\right)\right] -2 \left[W(z_j)-\left(g(z_{j+})+g(z_{j-})\right)\right]\\
        = 2 (L_{j+1}-L_j) = 2 \rmi (m_{j+1}-m_j)\in \rmi\mathbb{R} .
\end{eqnarray}
As a consequence, the coefficients $\lambda_j$ in~(\ref{idy}) are  given by
\begin{equation}
	\label{idy2}
	\lambda_j = \rmi r_j,\quad r_j = m_{j+1}-m_j  \in  \mathbb{R} .
\end{equation}
Furthermore, from~(\ref{mes}) we find that the $B$-periods are
 \begin{equation}
 	\label{ybs}
	B_i(\mathrm{y}(z)\rmd z)
	=
	- 4 \pi\rmi\sum_{j=1}^i \int_{\gamma_j} |\rmd z| \rho(z) \in \rmi\mathbb{R},
\end{equation}
and consequently~(\ref{idy}) implies
\begin{equation}
	\label{fina}
	\sum_{j=1}^{s-1} r_j  \im  B_i(\rmd \varphi_j)
	=
	\re \left(\sum_{n=0}^{N+1} t_n B_i(\rmd \Omega_n)\right).
 \end{equation}
Since the matrix of periods $\im B_i(\rmd \varphi_j)$ is positive definite~\cite{FA91},
the linear system~(\ref{fina}) uniquely determines the coefficients $r_j$ as functions
of the cut endpoints $\{a_k^{\pm}\}_{k=1}^s$ and the coefficients  $\{t_n\}_{n=1}^{N+1}$ of the potential $W(z)$. 
 
Therefore, we have the following method to find a system of equations for the cut endpoints:
\begin{description}
\item[(1)] We start with a function $y(z)$ of the form~(\ref{y1})--(\ref{y2}) and use the identity
\begin{equation}
 	\prod_{l=1}^{r} (z-\alpha_l)= \left(\frac{W'(z)}{w(z)}\right)_{\oplus}
 \end{equation}
to determine the double roots $\{\alpha_l\}_{l=1}^r$  of $y^2(z)$ in terms of the cut endpoints.
Then we express the coefficients of the polynomial $ y(z)^2-W'(z)^2$ in terms of the cut endpoints
and the coefficients of $W(z)$.
\item[(2)] From~(\ref{y1})--(\ref{ache})
			it is clear that
			\begin{eqnarray}
			\label{sh1}
			\nonumber
			y(z)^2-W'(z)^2
 			& =
			\left(\frac{W'(z)}{w(z)}\right)_{\ominus}
			\left[\left(\frac{W'(z)}{w(z)}\right)_{\ominus} w(z)^2-2 W'(z) w(z)\right]
			\\
			 & =
                         \mathcal{O}\left(z^{N+s-1}\right),\quad z\rightarrow \infty.
			\end{eqnarray}
			Moreover, in order to satisfy ~(\ref{sc2}), we impose that 
 			\begin{equation}
				\label{sh2}
				y(z)^2-W'(z)^2 = -4 z^{N-1} t_{N+1}+ \cdots.
			\end{equation}
			Therefore, we  equate to zero the coefficients of the powers $z^{N},\cdots, z^{N+s-1}$
			and to -$4 t_{N+1}$ the coefficient of $z^{N-1}$ in~(\ref{sh1}).
			Thus we obtain $s+1$ equations for the  $2 s$ cut endpoints.
\item[(3)] 		Finally, to obtain  $s-1$ additional equations, we express the differentials
			$\rmd\varphi_j$ and $\rmd\Omega_n$ in terms of the cut
			endpoints and solve the system~(\ref{fina}) to determine the unknowns $r_j$ as
			functions of the cut endpoints and of the coefficients of $W(z)$. 
			Then, in view of~(\ref{idy})--(\ref{idy2}) we impose
 			\begin{equation}
				\label{periods}
				\oint_{A_i} \mathrm{y}(z) \rmd z=2 \rmi r_i,\quad i=1,\ldots,s-1.
			\end{equation}
\end{description} 

There is an alternative and more intrinsic scheme for finding the cut endpoints using the expressions
of the Abelian differentials. Indeed, as a consequence of  the identities~(\ref{nho}),~(\ref{diff1}) and~(\ref{diff10})
of Appendix~A we have
\begin{equation}
	\label{eep}
	y(z) w(z) = 2 \sum_{n=0}^{N+1} t_n P_n(z) + 2 \rmi\sum_{i=1}^{s-1}r_i p_i(z).
\end{equation}
Hence if we set $z=a_j^{\pm}$ in this identity we find 
\begin{equation}
  \label{ce}
  \sum_{n=0}^{N+1} t_n P_n(a_j^{\pm})
  +
  \rmi \sum_{i=1}^{s-1}r_i p_i(a_j^{\pm})
  = 0,\quad j=1,\ldots,s.
\end{equation}
In particular for the hermitian case (see subsection 1.3)
\begin{equation}
	r_i = m_{i+1} - m_i =0,\quad i=1,\ldots,s-1,
\end{equation}
so that~(\ref{ce}) simplifies to
\begin{equation}
  \label{ceh}
  \sum_{n=0}^{N+1} t_n P_n(a_j^{\pm})=0,\quad j=1,\ldots,s,
\end{equation}
which is  the standard system used in hermitian random matrix models to determine
the asymptotic eigenvalue support~\cite{AL10}. In section~\ref{subsec:twocutcase} we will illustrate
for the usual cubic model how the new terms in the general equations~(\ref{ce}) are reduced
(via real parts and imaginary parts of periods of abelian differentials) to the calculation of standard
integrals, which in this particular case can be expressed in closed form in terms of elliptic functions.
\subsection{Construction of $S$-curves}
Once a solution of the endpoint equations has been obtained, the $y$-function~(\ref{y1}) is completely determined.
The next step is to find the cuts $\gamma_j$ connecting the respective pairs of cut endpoints $a_j^{\pm}$ and such that 
the $\rho(z)$ defined by~(\ref{mes}) is a normalized positive density and the $S$-property on
$\gamma=\gamma_1\cup \ldots\cup \gamma_s$ is satisfied. 

Let us define the function
\begin{equation}
	\label{G}
	G(z) = \int_{a_1^-}^z y(z'_+)  \rmd  z'.
\end{equation}
From~(\ref{cm}) and ~(\ref{y}) we have that  the $y$-function must satisfy
\begin{eqnarray}
	\label{idl}
	\nonumber
	\re \int_{a_1^-}^z y(z')  \rmd  z'
	& = \re  \left[(W(z)-2 g(z))-(W(a_1^-)-2 g(a_1^-))\right] \\
	& = U(z)-l, \quad z\in \mathbb{C}\setminus \gamma,
\end{eqnarray}
where $U(z)$ is the electrostatic potential~(\ref{loge}) and $l$ is some real constant.
Hence, in terms of $G(z)$ the equilibrium condition~(\ref{cm}) reads
\begin{equation}
	\label{cmp}
 	\re G(z)=0, \quad z\in \gamma.
\end{equation}
Note that different choices of the base point among the branch points $a_j^-$ in the integral~(\ref{G})
lead to conditions equivalent to~(\ref{cmp}).

Given a root $z_0$ of  $y^2(z)$ with multiplicity $m$, there are $m+2$ maximal connected components
(excluding any zeros of $y^2(z)$) of the level curve  
\begin{equation}
	\re \int_{z_0}^z y(z'_+)  \rmd  z'=0,
\end{equation}
which stem from $z_0$~\cite{SI95}.
These maximal components are called the Stokes lines outgoing from $z_0$ associated to the polynomial $y^2(z)$.
Stokes lines for a polynomial cannot make loops and end necessarily either at a different zero of $y(z)$
(lines of \emph{short} type) or at infinity (lines of \emph{leg} type). Therefore, the condition~(\ref{cmp}) means
that the cuts $\gamma_j$ must be short type lines  with cut endpoints $a_j^{\pm}$ of the polynomial $y^2(z)$.
It should be noticed that the function $y(z)$ is continuous on those short type lines which are not cuts.
In what follows we will denote by $\mathcal{X}_0$ the set of all the Stokes lines emerging from the simple
roots $a_j^{\pm}$ of $y(z)$ and by $\mathcal{X}$ the set of all the Stokes lines emerging from  all the roots of $y^2(z)$.
 
The positivity of the corresponding density~(\ref{mes}) also imposes that 
\begin{equation}
	\label{cmp2}
 	\im  G(z)>0, \quad z\in \gamma_j \setminus \{a_j^-,a_j^+\}, \; j=1,\ldots,s.
\end{equation}
However, the scheme of the above subsection implies
\begin{equation}
	y(z) = W'(z)-\frac{2}{z}+\mathcal{O}(z^{-2}), \quad z\rightarrow \infty,
 \end{equation}
so that~(\ref{norm}) holds. Therefore if~(\ref{cmp2}) is verified on $s-1$ cuts and the total charge on these cuts
is smaller than unity, then~(\ref{cmp2}) is also verified on the remaining cut.              

It is straightforward that if the cuts satisfy~(\ref{cmp}) and~(\ref{cmp2}) then the $S$-property is verified on $\gamma$,
and that we may characterize $S$-curves $\Gamma$ by imposing the following two additional conditions:
\begin{description}
	\item[(S1)] $\Gamma$ contains $\gamma$.
	\item[(S2)] $\Gamma$ does not cross any region of the complex plane where $\re  G(z)<0$ .
\end{description}
Indeed, as a consequence of~(S1) the path $\Gamma$ verifies the $S$-property with respect to the external potential $V(z)$.
Moreover, using~(\ref{idl}) we have that~(S2) implies the condition~(\ref{cm1}),
so that $\rho(z)$ is an equilibrium measure on $\Gamma$.

To implement condition~(S2) we need an explicit description of the set $\re G(z)>0$ in the complex plane.
It is helpful to observe that points in the neighborhood of a cut satisfy  $\re  G(z)<0$, while the remaining connected
lines of the level set $\re  G(z)=0$ separate regions where $\re  G(z)<0$ from regions where $\re G(z)>0$.
These properties can be proved as follows.  From~(\ref{G}) we have that the derivatives of $\re G$ with respect
to the cartesian coordinates are
\begin{equation}
   \label{dG}
    \frac{\partial}{\partial x} \re  G(z) =  \re   y(z),
    \quad
    \frac{\partial}{\partial y} \re  G(z)=- \im   y(z).
\end{equation}
Then take for instance a point $z_++\delta z$ near to a point $z_+$ of a cut and to the left of the cut
(i.e. $\delta z=\rmi\,\rmd z$).  Then since $\re G(z_+)=0$ and using~(\ref{dG}) we have
\begin{equation}
\re G(z_++\delta z)\simeq -\im \left(y(z_+) \rmd z \right)=-2 \pi \rho(z) |\rmd z|<0.
\end{equation}
The same result is obtained for points $z_-+\delta z$ near to a point $z_-$ of a cut and to the right of the cut
(i.e. $\delta z=-\rmi\,\rmd z$) taking into account that  $y(z_-)=-y(z_+)$. The corresponding statement for the other
connected lines  verifying $\re  G(z)=0$ follows similarly using the continuity of $y(z)$ on them. 

Equation~(\ref{dG}) also shows that if a Stokes line $c$ emerging  from one cut endpoint $a_j^{\pm}$
meets a zero $z_\mathrm{c}$ of $y(z)$ different from  $a_j^{-}$ and $a_j^{+}$
both partial derivatives of the curve $\re G(x,y)=0$ vanish at $z_\mathrm{c}$ and therefore $c$ has a critical point at $z_\mathrm{c}$.
These situations arise in particular at phase transitions of equilibrium densities  in which the number of cuts changes.
\subsection{The one-cut case\label{sec:1cut}}
In the one-cut case we will drop the general notation and denote the cut endpoints by
$a_1^{-}= a$ and $a_1^{+}= b$ respectively. The scheme of section~\ref{sec:eqendpts} to determine the
cut endpoints reduces to identifying the coefficients of $z^{N-1}$ and $z^N$ in~(\ref{sh1}), where 
\begin{equation}
    y(z) = \left(\frac{W'(z)}{w(z)}\right)_{\oplus}  w(z),\quad w(z)=\sqrt{(z-a)(z-b)}.
\end{equation}
The resulting equations for $a$ and $b$ are often  simpler when expressed in terms of 
\begin{equation}
	\beta = \frac{a+b}{2}, \quad \delta = \frac{b-a}{2}.
\end{equation}
Moreover, in this case  the function 
\begin{equation}
	G(z) = \int_{a}^z y(z'_+)  \rmd  z',
\end{equation}
is  given by 
\begin{equation}
	\label{Ge}
	G(z) = \left(\frac{W(z)}{w(z)}\right)_{\oplus} w(z)- \log\left(\frac{z-\beta+w(z)}{a-b}\right)^2-\log 4.
\end{equation}
To prove this identity we recall the form of the function $y(z)=h(z)  w(z)$ and look for a decomposition 
\begin{equation}
	G(z) = Q(z) w(z) + c \int_{a}^z \frac{\rmd z'}{w(z')},
\end{equation}
where $Q(z)$ is a polynomial and $c$ a complex constant. Differentiating this equation with respect
to $z$ and multiplying by $w(z)$ we get
\begin{eqnarray}
	h(z) w(z)^2
	&=
	\left(W'(z)-2 g'(z)\right) w(z)
	=
	\left(Q(z) w(z)\right)' w(z)+c\nonumber\\
	&=
	\left(Q(z) w(z)-f(z)\right)' w(z),
\end{eqnarray}
with
\begin{equation}
	f'(z)=-\frac{c}{w(z)}.
\end{equation}
Hence
\begin{equation}
	f(z) = -c \log\left(z-\beta+w(z)\right),
\end{equation}
and
\begin{equation}
	Q(z) = \frac{W(z)}{w(z)}+\frac{f(z)-2 g(z)}{w(z)}+\frac{C}{w(z)},
\end{equation}
for a certain complex constant $C$. Since $Q(z)$ is a polynomial, the logarithmic terms in $f(z)-2 g(z)$ must cancel,
and taking into account that
\begin{equation}
	g(z) = \log z + \mathcal{O}(1/z),\quad z\rightarrow \infty,
\end{equation}
we get that $c=-2$ and
\begin{equation}
	Q(z) = \left(\frac{W(z)}{w(z)}\right)_{\oplus}.
\end{equation}
\subsection{The Gaussian model}
A simple illustration of the above method is provided by the Gaussian model
\begin{equation}
	\label{gau}
	W(z) = \frac{z^2}{2}.
\end{equation}
In this case only spectral curves with one cut may arise. Moreover, $y(z)=\sqrt{(z-a)(z-b)}$  and $f(z)=-4$.
Then~(\ref{c1b}) leads to
\begin{equation}
	b = -a = 2.
\end{equation}
If we take the cut  $\gamma$  as the interval $[-2,2]$  then
\begin{equation}
	y(z_+) = \rmi  |z^2-4|^{1/2},\quad z\in \gamma
\end{equation}
and $\gamma$ satisfies the $S$-property.

\begin{figure}
    \begin{center}
        \includegraphics{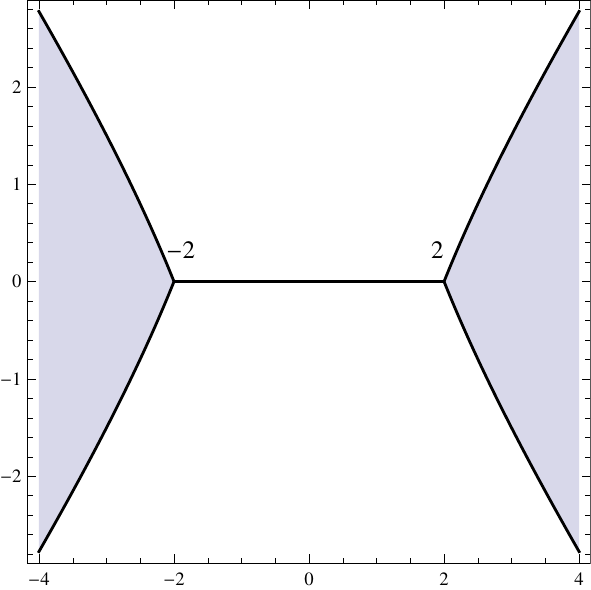}
    \end{center}
    \caption{The set $\mathcal{X}_0$ and the regions $ \re G(z)>0$ (shadowed regions) for the Gaussian model.\label{fig:ga}}
\end{figure}

Figure~\ref{fig:ga} shows the Stokes lines emerging from the cut endpoints of the Gaussian model as well as the
set $ \re G(z)>0$ where the cut may be continued into an $S$-curve. A possible choice is $\Gamma=\mathbb{R}$.
Then we may define $\log(z-z')$  as the principal branch of the logarithm and we have
\begin{equation}
  \log(z_+-z')+\log(z_--z')=2\log |z-z'|,\quad z,z'\in \mathbb{R}.
\end{equation}
\section{The cubic model\label{sec:tcm}}
We will now discuss the cubic model 
\begin{equation}
    \label{eq:w3}
    W(z) = \frac{z^3}{3} - t z,
\end{equation}
where $t$ is an arbitrary complex number. For $t=0$ the  model has been rigorously  studied by Dea\~{n}o, Huybrechs
and Kuijlaars~\cite{DE10}. Recent results for $t\in \mathbb{R}$ have been communicated by Lejon~\cite{LE12}.
The phase structure of the corresponding random matrix model  has been studied by David~\cite{DA91}
and Mari\~{n}o~\cite{MAR10}.
\subsection{The one-cut case}
Using the notation specific for the one-cut case introduced in section~\ref{sec:1cut} we have
\begin{equation}
    \label{eq:w3y1}
    y(z) = \left(z+\beta\right) \sqrt{(z-\beta)^2-\delta^2},\quad f(z)=-4 z+b_0,
\end{equation}
and~(\ref{sh1}) and~(\ref{sh2}) lead to the following system of equations for the cut endpoints:
\begin{equation}
    \label{eq:y1h1}
    2 \beta^2 + \delta^2 = 2 t,
\end{equation}
\begin{equation}
    \label{eq:y1h2}
    \beta \delta^2 = 2 .
\end{equation}
Therefore $\beta$  satisfies the cubic equation
\begin{equation}
    \label{eq:beta}
    \beta^3 - t \beta + 1 = 0
\end{equation}
and $\delta$ is determined by
\begin{equation}
    \label{eq:delta1}
    \delta^2 = \frac{2 }{\beta}.
    \end{equation}
The cubic equation~(\ref{eq:beta}) defines a three-sheeted Riemann surface $\Xi$  of genus zero for
$\beta$ as a function of $t$. The function  $\beta(t)$ is determined in terms of three branches
\begin{equation}
    \label{eq:betak}
    \beta_k(t) = - \frac{t}{3 \Delta_k} - \Delta_k, \quad (k=0,1,2)
\end{equation}
where
\begin{equation}
    \Delta_k = \rme^{\rmi 2\pi k/3}
                     \sqrt[3]{\frac{1}{2} + \sqrt{\frac{1}{4} - \left(\frac{t}{3}\right)^3}}
\end{equation}
and where the roots take their respective principal values. There are three finite branch points 
\begin{equation}
    t^{(k)} =  \frac{3}{2^{2/3}} \rme^{\rmi 2\pi k/3}, \quad (k=0,1,2)
\end{equation}
at which $\beta_1(t^{(0)})=\beta_2(t^{(0)})$, $\beta_0(t^{(1)})=\beta_1(t^{(1)})$,
and $\beta_0(t^{(2)})=\beta_2(t^{(2)})$ respectively.

\begin{figure}
    \begin{center}
        \includegraphics[width=15cm]{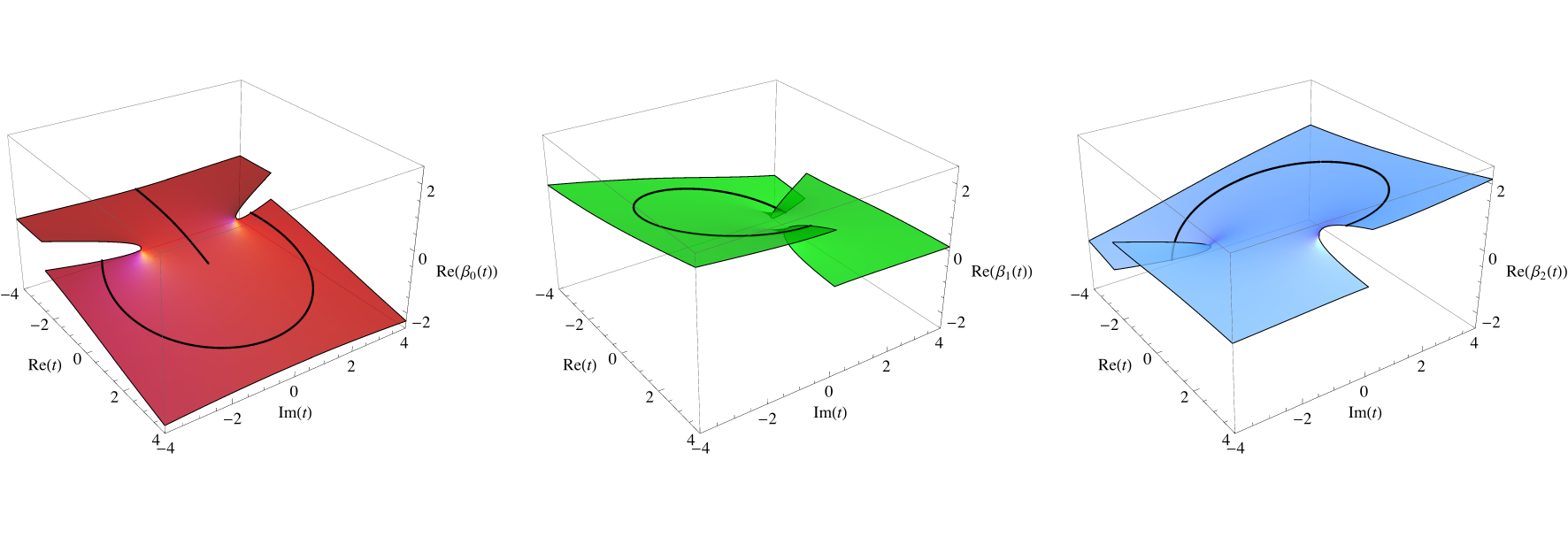}
    \end{center}
    \caption{Real parts of the branches of $\beta(t)$.\label{fig:rs}}
\end{figure}

In the three separate plots of figure~\ref{fig:rs} we show the real parts of the three branches
of the Riemann surface~(\ref{eq:beta}). As an aid to guide the eye, we also plot two paths on the surface.
The first path starts at the origin $t=0$ in $\beta_0(t)$ (i.e., at $\beta_0(0)=-1$)
and proceeds to the left without leaving this sheet.
The second path corresponds to $|t|=3$ (larger than the modulus of the branch points $t^{(k)}$):
note that the path stays in the branch $\beta_0(t)$ from the real axis $\arg t = 0$ to $\arg t = 2\pi/3$,
proceeds to the $\beta_1(t)$ branch from $\arg t = 2\pi/3$ to $\arg t = 2\pi$,
then to $\beta_2(t)$ from $\arg t=2\pi$ to $\arg t = 10\pi/3$, and back to the branch $\beta_0(t)$
form $\arg t = 10\pi/3$ to the real axis $\arg t = 4\pi$.
\begin{figure}
    \begin{center}
        \includegraphics{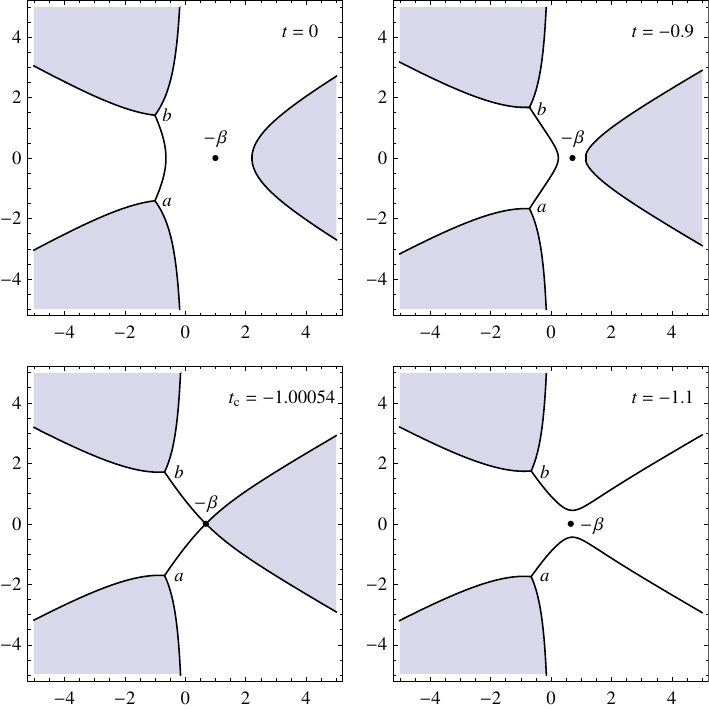}
    \end{center}
    \caption{Sets $\mathcal{X}_0$ of Stokes lines  calculated according  to the branch $\beta_0(t)$  starting at $t=0$
    and proceeding along the negative $t$ axis (path to the left in figure~\ref{fig:rs}). The shaded areas are regions with $\re G(z)>0$.
    \label{fig:pathneg}}
\end{figure}

Figure~\ref{fig:pathneg} shows the sets $\mathcal{X}_0$ of all the Stokes lines of the roots $a,b$ 
of the function $y^2(z)$ corresponding to $\beta_0(t)$ for negative real values of $t$. The path on $\beta_0(t)$ 
starts at $t=0$ and proceeds along the negative $t$ axis (the path to the left in figure~\ref{fig:rs}). Note the two simple
zeros $a$ and $b$, each one with three Stokes lines stemming at equal angles of $2\pi/3$. In the two first plots, corresponding
to $t=0$ and $t=-0.9$, we find a short  connecting $a$ and $b$, so that we get a cut  satisfying the $S$-property. However, for a
critical value $t_\mathrm{c}\approx -1.000 54$ the double zero of $y^2(z)$  meets this cut giving rise to a singular curve,
and beyond that point  there is no Stokes line joining $a$ to $b$. This indicates that for $t<t_\mathrm{c}$ the branch $\beta_0(t)$
does not lead to a cut satisfying the $S$-property. (This interpretation is in agreement with the main theorem in~\cite{LE12}.)

\begin{figure}
    \begin{center}
        \includegraphics{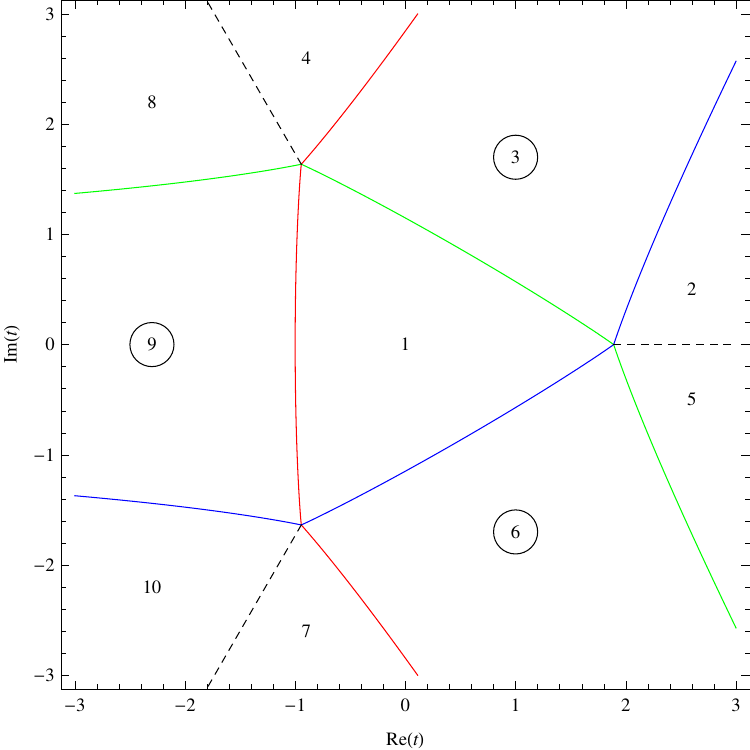}
    \end{center}
    \caption{The solid lines represent the solutions of~(\ref{eq:gabc}) for  $k=0,1,2$, with colors matching those
                 of the respective branches in figure~\ref{fig:rs}.
                 The dashed lines are the cuts ($|t|>3/2^{2/3},\;\arg t= 0, \pm 2\pi/3$) of the Riemann surface~(\ref{eq:beta}).\label{fig:ph}}
\end{figure}

In fact, we can find an analytic condition (which, however, has to be solved numerically) for the set of complex values of $t$
such that  $-\beta\in\mathcal{X}_0 $, where $\mathcal{X}_0$  is the set of Stokes lines of  $a$ and $b$.
Using~(\ref{Ge}) we find that the $G$ functions corresponding to the branches $\beta_k(z)$  are
\begin{eqnarray}
	G_k(z)
	&=& \frac{1}{3} \sqrt{(z- \beta_k)^2-\delta_k ^2}  \left(z^2+ \beta_k z+\beta_k^2-3 t+\frac{\delta_k ^2}{2}\right)
	\nonumber\\
  	& &{}-\log\left(\frac{\beta_k-z-\sqrt{(z- \beta_k)^2-\delta_k ^2}}{\delta_k }\right)^2 .                 
\end{eqnarray}
Hence the  condition for $-\beta_k\in\mathcal{X}_0 $ is
\begin{equation}
    \label{eq:gabc}
    \re G_k(-\beta_k(t)) = 0,
\end{equation}
where
\begin{equation}
    \label{eq:gab}
    G_k(-\beta_k) = -\frac{1}{3} \sqrt{4 \beta_k^2-\delta_k ^2} \left(2 \beta_k^2+\delta_k ^2\right)
                              -\log\left(\frac{2 \beta_k-\sqrt{4 \beta_k ^2-\delta_k^2}}{\delta_k }\right)^2.
\end{equation}
In figure~\ref{fig:ph} we show the curves in the complex $t$-plane determined by the solutions of~(\ref{eq:gabc}),
with colors matching those of the corresponding branches in figure~\ref{fig:rs}. In addition each region has been identified
with a number that will be used in our forthcoming discussion of the phase structure.
\subsection{The two-cut case\label{subsec:twocutcase}}
In the two-cut case we will denote $a_1^{-}=a$, $a_1^{+}=b$, $a_2^{-}=c$, $a_2^{+}=d$ and $r_1=r$.
Now we have
\begin{equation}
    \label{eq:w3y2}
    y(z) = \sqrt{(z-a)(z-b)(z-c)(z-d)},\quad f(z)=-4  z+b_0,
\end{equation}
and~(\ref{sh1}), (\ref{sh2}) and~(\ref{periods}) lead the following system of equations for the four cut endpoints
\begin{eqnarray}
	\label{eq:y2h1}
	a b c + a b d + a c d + b c d =  4 , \\
	\label{eq:y2h2}
	a b + a c + b c + a d + b d + c d  =  - 2 t, \\
	\label{eq:y2h3}
	a + b + c + d =  0, \\
	\label{eq:y2h4}
	\int_b^c y(z_+) \rmd z  =  \rmi r.
\end{eqnarray}
We recall that $r$ is given in terms of $B$-periods
\begin{equation}
	B(\rmd \omega)=\oint_{B}\rmd \omega=-2 \int_a^b \rmd \omega
\end{equation}
 (we drop the subindex, i.e., $B=B_1$) by equation~(\ref{fina})
\begin{equation}\label{monster}
	r = \frac{\re B\left(\frac{1}{3} \rmd \Omega_3-t \rmd \Omega_1-\rmd \Omega_0\right)}{\im B(\rmd \varphi)}.
\end{equation}
Taking into account that~(\ref{eq:y2h1})--(\ref{eq:y2h3}) imply
\begin{equation}
	y(z) = z^2-t-\frac{2}{z}+\cdots,\quad z\rightarrow\infty,
\end{equation}
it follows that 
\begin{equation}
	\label{monster2}
	r = \frac{\re \left(\mathcal{B}_4-2 t \mathcal{B}_2-4 \mathcal{B}_1+C \mathcal{B}_0\right)}{\im \left(\mathcal{B}_0/\mathcal{A}_0\right)},
\end{equation}
where $\mathcal{A}_n$, $\mathcal{B}_n$ denote the integrals
\begin{equation}
	\mathcal{A}_n=\int_b^c \frac{z^n}{y(z_+)} \rmd z,\quad \mathcal{B}_n=\int_a^b \frac{z^n}{y(z_+)} \rmd z,
\end{equation}
and 
\begin{equation}
	C=\frac{-\mathcal{A}_4+2 t \mathcal{A}_2+4 \mathcal{A}_1}{\mathcal{A}_0}.
\end{equation}

It is clear that in general the system~(\ref{eq:y2h1})--(\ref{eq:y2h4}) must be solved numerically.  
But even so, it would be very difficult to attempt a direct numerical solution without a well identified
initial approximation. However, we can take advantage of our knowledge
of the critical curves~(\ref{eq:gabc}) and the corresponding explicit solutions for the one-cut endpoints
given by~(\ref{eq:betak}), and proceed iteratively by small increments in $t$ using as initial approximation
at each step the results of the previous one. Once the cut endpoints $a$, $b$, $c$ and $d$ for a certain value
of $t$ have been calculated, the corresponding Stokes lines are also calculated numerically.

\begin{figure}
    \begin{center}
        \includegraphics{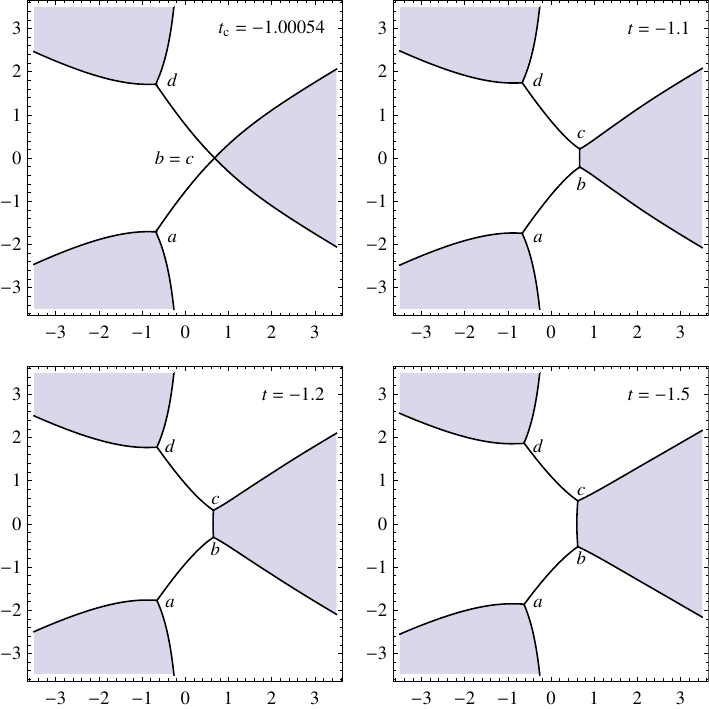}
    \end{center}
    \caption{Splitting of  a cut.\label{fig:1-10}}
\end{figure}

\begin{figure}
    \begin{center}
        \includegraphics{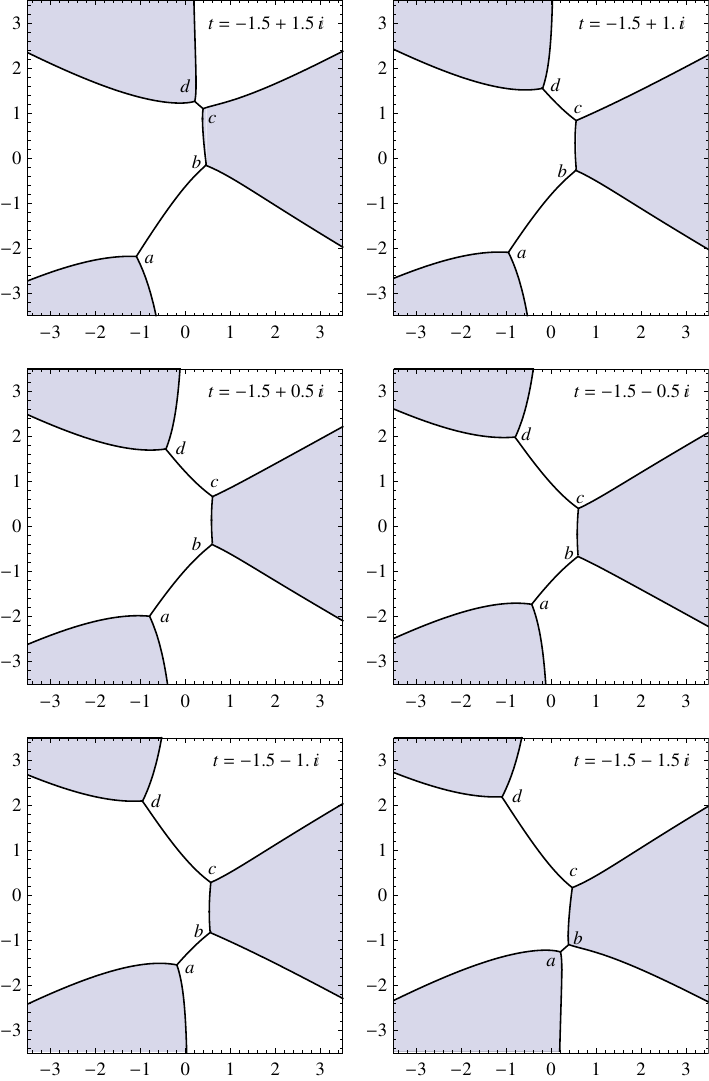}
    \end{center}
    \caption{Birth and death of  a cut at a distance .\label{fig:1-9}}
\end{figure}

Figures~\ref{fig:1-10} and~\ref{fig:1-9}  show the sets $\mathcal{X}_0=\mathcal{X}$ of all the Stokes lines
stemming from the simple roots  $a$, $b$, $c$ and $d$ for values of $t$ crossing critical lines of figure~\ref{fig:ph}.
In figure~\ref{fig:1-10} we proceed along the negative $t$ axis beyond the critical value $t_\mathrm{c}\approx -1.000 54$
(i.e., to the part of the path corresponding to $t < -1.000 54$ in the first graph of figure~\ref{fig:rs})
and we find a ``splitting of a cut'' at the crossing from region~1 to region~9 in figure~\ref{fig:ph},
in agreement with the theoretical result of~\cite{LE12}. In figure~\ref{fig:1-9}  we have crossed vertically
from region~8 into region~9,
and find a process of ``birth of a cut at a distance'' with cut endpoints $c$ and $d$; the graph corresponding to $t=-1.5$,
not shown in the figure, is precisely the last graph in figure~\ref{fig:1-10}; and as we proceed
further down from region~9 to region~10 we find the symmetric ``death of a cut at a distance'' with cut endpoints $a$ and $b$.
In the next section these interpretations are confirmed by numerical calculations of zeros of orthogonal polynomials.
\subsection{Asymptotic zero distributions of orthogonal polynomials\label{sec:phst}}
As we discussed in section~\ref{sec:eds}, to determine the asymptotic  zero distribution
of a given  family of orthogonal polynomials~(\ref{pol1}) on a path $\Gamma$,
we must find an $S$-curve  in the same homology class as $\Gamma$  and connecting
the same pair of convergence sectors at infinity. Then the desired zero counting measure is the equilibrium
measure on the $S$-curve.

\begin{figure}
    \begin{center}
        \includegraphics{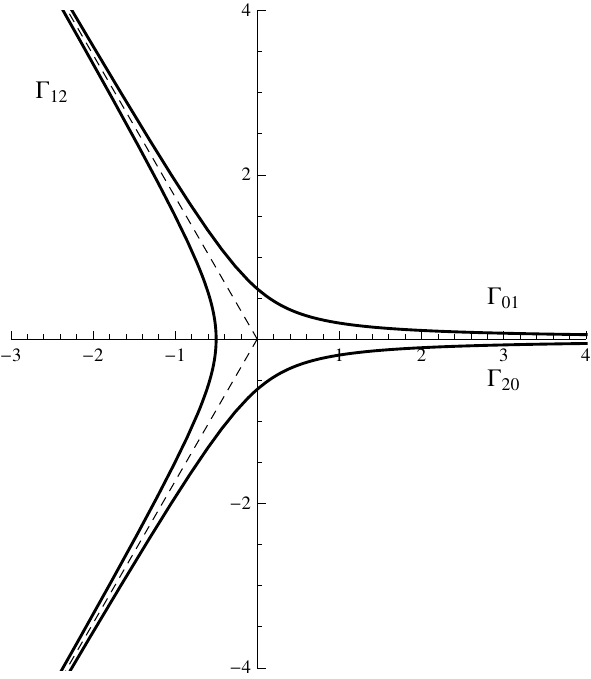}
    \end{center}
    \caption{Infinite simple curves connecting convergence sectors of the complex cubic model.\label{fig:3n}}
\end{figure}

The cubic exponential weight $\exp(-n(z^3/3-t z))$ decays in three sectors $S_k$ of opening $\pi/3$
of the complex $z$ plane centered around the rays $\lambda_k =\{z: \arg z=2 \pi k/3\}$,  $k=0,1,2$.
Let us denote by $\Gamma_{ij}$ $(i\neq j)$  simple paths with asymptotic directions $\lambda_i$ and
$\lambda_j$ as indicated in figure~\ref{fig:3n} and, for concreteness, consider the problem of determining
$S$-curves $\Gamma$ in the same homology class and with the same asymptotic directions of $\Gamma_{12}$.
The graph corresponding to $t=0$ in figure~\ref{fig:pathneg} shows that the cut $ab$ can be prolonged
both upwards and downwards into the shaded regions which contain the asymptotic directions $\lambda_1$
and $\lambda_2$ respectively, and therefore into a full $S$-curve homologous to $\Gamma_{12}$.
This is no longer true for $t<t_\mathrm{c}$, as the graph corresponding to $t=-1.1$ in figure~\ref{fig:pathneg}
shows: in fact, the cut $ab$ has disappeared. However, the graph for $t=-1.1$ in figure~\ref{fig:1-10}
features the two cuts $ab$ and $cd$, which can be prolonged into the same sectors via the shaded region
in the right part of the figure. Therefore, for this value of $t$ we have a full two-cut  $S$-curve.

This type of analysis which combines the theoretical results of section~\ref{sec:ced} with numerical calculations
show that in the case of $\Gamma_{12}$ the branches $\beta_0$, $\beta_1$ and $\beta_2$ can be used
to generate a one-cut  $S$-curve for the cubic model when $t$ is in the regions~1 to~7, 8 and~10 of  figure~\ref{fig:ph},
respectively. For $\Gamma_{12}$ the encircled region~9 represents the two-cut region.
Similar (symmetric) situations arise for the cases of $\Gamma_{01}$ and $\Gamma_{20}$, for which the two-cut
regions are the encircled regions~3 and~6 respectively.

\begin{figure}
    \begin{center}
        \includegraphics{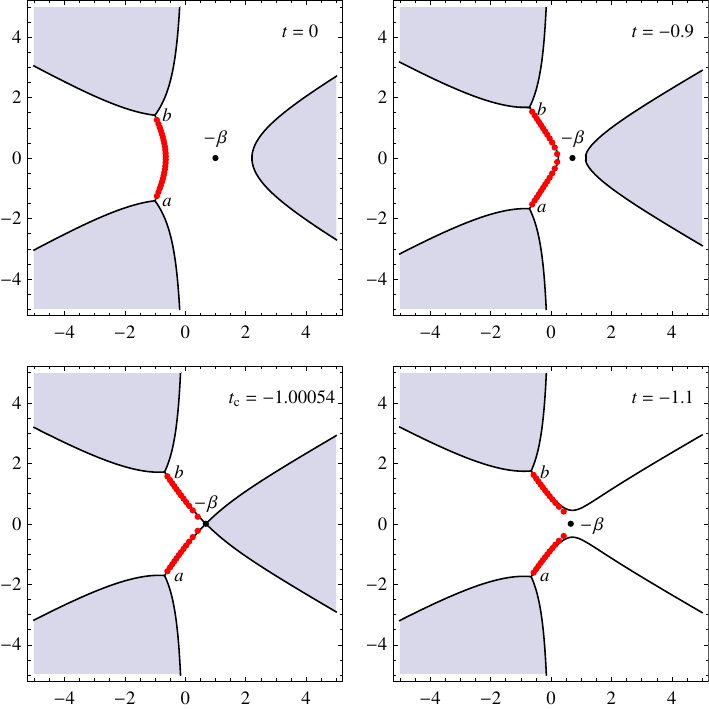}
    \end{center}
    \caption{Zeros of $p_{24}(z)$ superimposed to the splitting of  a cut in figure~\ref{fig:pathneg}.\label{fig:zd1}}
\end{figure}

\begin{figure}
    \begin{center}
        \includegraphics{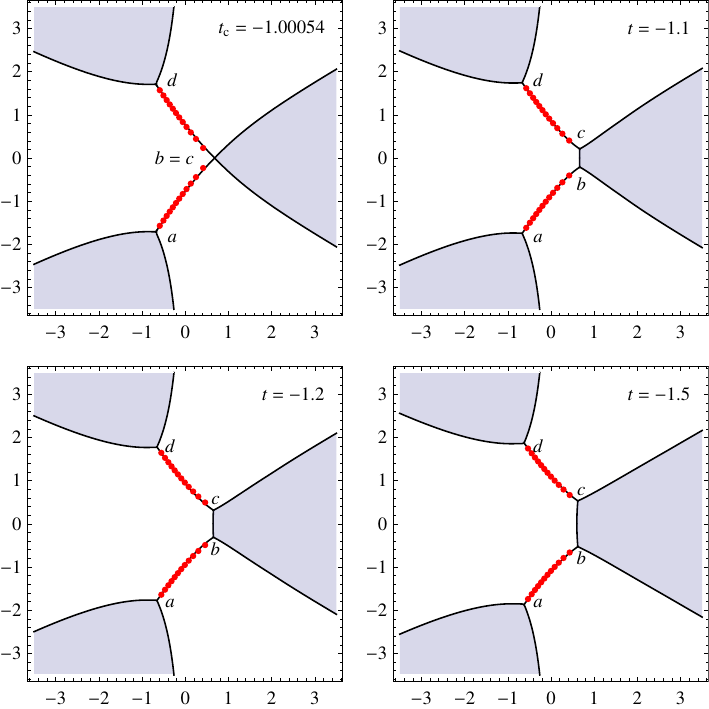}
    \end{center}
    \caption{Zeros of $p_{24}(z)$ superimposed to the splitting of  a cut in figure~\ref{fig:1-10}.\label{fig:zd2}}
\end{figure}

\begin{figure}
    \begin{center}
        \includegraphics{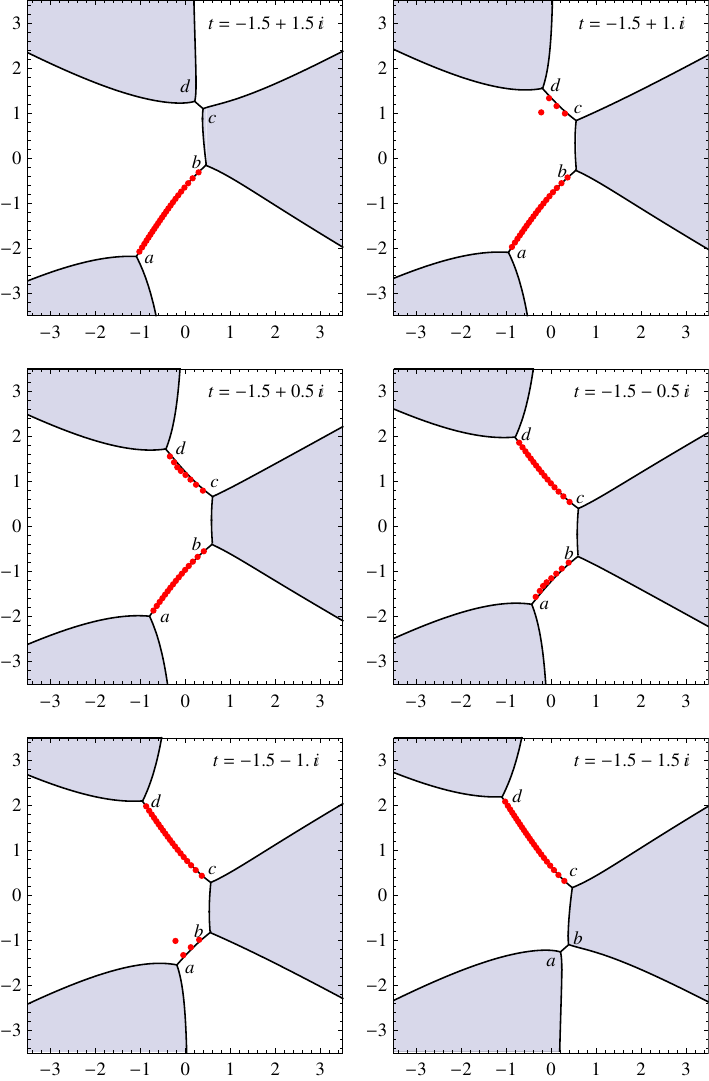}
    \end{center}
    \caption{Zeros of $p_{24}(z)$ superimposed to the birth and death of a cut at a distance in figure~\ref{fig:1-9}.\label{fig:zd3}}
\end{figure}

As a check of the consistency of these results with Theorem~1, in figures~\ref{fig:zd1},~\ref{fig:zd2} and~\ref{fig:zd3}
we superimpose to the graphs of figures~\ref{fig:pathneg},~\ref{fig:1-10} and~\ref{fig:1-9} the zeros of the
corresponding polynomials $p_{n}(z)$ with degree $n=24$,
which we have generated by recurrence formulas to minimize numerical
errors. In figures~\ref{fig:zd1} and~\ref{fig:zd2}, which exemplify the splitting of a cut, as $t$ decreases along the
negative real axis and due to the symmetry of the situation, the 24 zeros split evenly into the two sets of 12 zeros
following closely the positions of the cuts that correspond to the limit $n\to\infty$. In figure~\ref{fig:zd3},
which exemplifies the birth and death of a cut at a distance, what we find numerically as the value of $t$
descends vertically from $t=-1.5+1.5\rmi $ to $t=-1.5-1.5\rmi $ is that all the 24 zeros
lie initially on the lower cut $ab$, and start travelling upwards one by one, thus populating the upper cut $cd$
and depopulating the lower $ab$. This behavior is particularly clear in the second graph
(corresponding to $t=-1.5+\rmi$), in which the fourth zero is ``arriving'' at the upper cut, and in the symmetric
graph (corresponding to $t=-1.5-\rmi$), in which the 21st zero is ``leaving'' the lower cut.

A phase diagram for the cubic random matrix model with a two-cut region with the same shape as region~3
in figure~\ref{fig:ph} was presented in~\cite{DA91}.  It is also worth noticing that in terms of the variable $w=t^{-3/2}$ 
the curve~(\ref{eq:gabc}) looks quite similar to the genus~0 \emph{breaking curve} found in~\cite{BE12}
for the family of orthogonal polynomials associated to the quartic potential
\begin{equation}
	W(z) = \frac{w}{4}z^4 + \frac{1}{2}z^2.
\end{equation}
However, the curve in~\cite{BE12} is only symmetric with respect the real $w$-axis, while the curve for the cubic model
is symmetric with respect both real and imaginary axes. Another important difference between the curve for the cubic model
and that for the quartic model  is that for this later  there exist
genus~0 and genus~1 \emph{breaking curves} (see figure~4 in~\cite{BE12}), although implicit equations
for genus~1 curves are provided only for the symmetric case~\cite{BE12}.
\section{Generalizations and concluding remarks\label{sec:gcr}}
A generalization of the $S$-property~(\ref{s20}) arises in the study of dualities between supersymmetric
gauge theories and string models on local Calabi-Yau manifolds $\mathcal{M}$
of the form~\cite{CA01,DI02,DI022,DI95,SE94,BE95,CA03}
\begin{equation}
	\label{cy}
	W'(z)^2+f(z)+u^2+v^2+w^2=0,
\end{equation}
where $W(z)$ and $f(z)$ are polynomials such that $ \deg   f= \deg   W-2$. The manifold $\mathcal{M}$
can be regarded as a fibration of two-dimensional complex spheres on the spectral curve $y^2=W'(z)^2+f(z)$.
Most of  the string model information encoded in $\mathcal{M}$ can be described in terms of the  spectral curve,
and its associated complex density~(\ref{mes}).   
These spectral curves satisfy the condition~(\ref{s20}) for the $S$-property,  but they do not determine
an equilibrium density since~(\ref{mes}) provides in general  a complex density.
As a consequence the complex electrostatic potential is locally constant on the support of $\rho(z)$
\begin{equation}
	\label{s2n}
	\mathcal{U}(z)=L_j, \quad  z\in \gamma_j,\quad j=1,\ldots, s,
\end{equation}
but the real parts of the constants $L_j$ are, in general, different.
In this case the cut endpoints are determined by~(\ref{sh1}),~(\ref{sh2}) and, instead of~(\ref{periods}),
by the constraints
\begin{equation}
	\label{ss}
	\int_{\gamma_j} \rho(z)  |\rmd z| = S_j,
\end{equation}
where $\rho(z)$ is the complex density~(\ref{mes}) and $S_j$ are a given set of nonzero complex values
('t~Hooft parameters). Finally, instead of the single quadratic differential $y^2(z) (\rmd z)^2$, in this case $s$
in general different quadratic differentials $\rme^{-\rmi 2\arg S_j} y^2(z) (\rmd z)^2 $ are required to determine
the cuts $\gamma_j$ as Stokes lines
\begin{equation}
	 \re \left(\rme^{-\rmi\arg S_j} \int_{a_j^{-}}^z y(z)  \rmd  z\right)=0,\quad z\in \gamma_j. 
\end{equation}
We believe that these more general spectral curves can be characterized and classified
using an analysis similar to that of the present paper.
\section*{Acknowledgments}
We thank Prof.~A.~Mart\'{\i}nez  Finkelshtein for useful conversations and for calling
our attention to the work~\cite{RA11}. The financial support of the Ministerio de Ciencia e Innovaci\'on
under projects FIS2008-00200 and FIS2011-22566 is gratefully acknowledged.
\section*{Appendix A}
In this appendix we briefly discuss the elements of the  theory of Abelian differentials in Riemann surfaces that we use
in section~\ref{sec:eqendpts}.

\begin{figure}
    \begin{center}
        \includegraphics{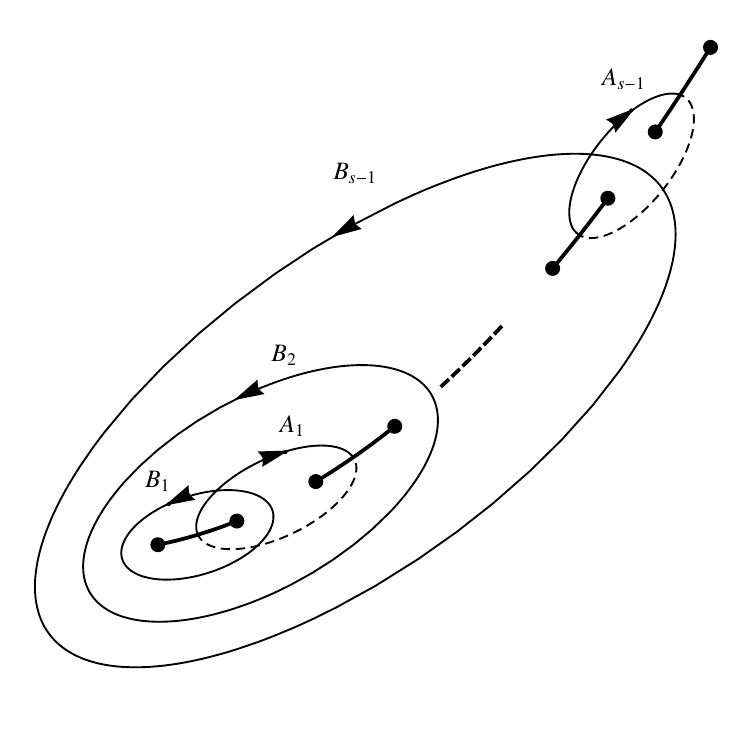}
    \end{center}
    \caption{Homology basis.\label{fig:hb}}
\end{figure}

Let us denote by $M$ the hyperelliptic Riemann surface  associated to the curve~(\ref{1.1}).
The two branches $w_1(z)$ and $w_1(z)=-w_2(z)=w(z)$  characterize $M$ as a double-sheeted
covering of the extended complex plane:
\begin{equation}
	M = M_1\cup M_2,\quad M_i = \{Q=(w_i(z),z)\}.
\end{equation}
The homology basis $\{A_i, B_i\}_{i=1}^{s-1}$ of cycles in $M$ is defined as shown in figure~\ref{fig:hb},
and the corresponding periods of a differential $\rmd \omega$ in $M$ will be denoted by
\begin{equation}
	\label{notp}
  	A_i(\rmd \omega) = \oint_{A_i} \rmd \omega,
	\quad
	B_i(\rmd \omega)=\oint_{B_i} \rmd \omega.
 \end{equation}
We  introduce the following Abelian differentials in $M$:
 \begin{description}
	\item[(1)] The canonical basis of first kind (i.e., holomorphic) Abelian differentials  $\{ \rmd \varphi_i\}_{i=1}^{s-1}$
			with the normalization $A_i(\rmd \varphi_j )=\delta_{ij}$. These differentials can be written as
			\begin{equation}
				\label{nho}
   				\rmd \varphi_j(z) = \frac{p_j(z)}{w(z)}\rmd z,
			\end{equation}
			where the $p_j(z)$ are polynomials of degree not greater than $s-2$ uniquely determined
			by the normalization conditions.
	\item[(2)] The second kind Abelian differentials $\rmd \Omega_k$  $(k\geq 1)$ whose only poles are at
			$\infty_1$, such that
			\begin{equation}
				\label{ok}
				\rmd \Omega_k(Q)
				=
				(k z^{k-1}+\mathcal{O}(z^{-2}))\rmd z,\quad Q\rightarrow \infty_1,\quad z=z(Q),
			\end{equation}
			and normalization $A_i(\rmd \Omega_k)=0$ $(i=1,\ldots,s-1)$. It is easy to see that
			\begin{equation}
 				\label{diff1}
				\rmd \Omega_k=\left(\frac{k}{2} z^{k-1}+\frac{P_k(z)}{w(z)}\right)\rmd z,
			\end{equation}
			where the $P_k(z)$ are polynomials of the form
			\begin{equation}
				\label{ek}
  				P_k(z) = \frac{k}{2} (z^{k-1} w(z))_{\oplus}+\sum_{i=0}^{s-2}c_{k i} z^i
			\end{equation}
			and the coefficients $c_{k i}$ are uniquely determined by the normalization conditions.
	\item[(3)] The third kind Abelian differential  $\rmd\Omega_0$ whose only poles are at $\infty_1$ and $\infty_2$, such that
			\begin{equation}
  				\label{o0}
				\rmd \Omega_0(Q)
				=
				\left\{\begin{array}{ll}
                                        \displaystyle\left(\frac{1}{z}+\mathcal{O}(z^{-2})\right)\rmd z,&  Q\rightarrow \infty_1\\
                                       \displaystyle\left(-\frac{1}{z}+\mathcal{O}(z^{-2})\right)\rmd z,& Q\rightarrow \infty_2,
                                \end{array}\right.\quad z=z(Q).
			\end{equation}
			and normalization $A_i(\rmd \Omega_0)=0$ for all $i=1,\ldots,s-1$. It follows that
			\begin{equation}
				\label{diff10}
				\rmd\Omega_0 = \frac{P_0(z)}{w(z)} \rmd z,
			\end{equation}
			where $P_0(z)$ is a polynomial of the form
			\begin{equation}
				\label{e0}
  				P_0(z)=(z^{-1} w(z))_{\oplus}+\sum_{i=0}^{s-2}c_{0 i} z^i
			\end{equation}
			and the coefficients $c_{0i}$ are uniquely determined by the normalization conditions.
\end{description}
For instance, in the one-cut case ($s=1$) we have
\begin{equation}
  \label{q1}
  P_k(z) = \left(\delta_{k0}+\frac{k}{2}\right)\left(z^{k-1} \sqrt{(z-a)(z-b)}\right)_{\oplus},
\end{equation}
and the first three polynomials are
\begin{eqnarray}
	\label{zero}
	P_0(z)=1,\\
	P_1(z)=\frac{1}{2} z-\frac{1}{4} (a+b),\\
	P_2(z)=z^2-\frac{1}{2} (a+b) z-\frac{1}{8} (a-b)^2.
\end{eqnarray}
\section*{References}
\providecommand{\newblock}{}

\end{document}